\documentstyle[12pt,fleqn]{article}
\begin{document}
\renewcommand{\thefootnote}{\fnsymbol{footnote}}
\setcounter{page}{1}
\begin{titlepage}
\begin{flushright}
\large
TSU-HEPI 94-04 \\
May 1994
\end{flushright}
\vspace{3.cm}
\begin{center}
{\Large\bf Estimating Condensates From the Difference}\\
{\Large\bf of $\tau$ Branching Fractions}\\
\vspace{1.0cm}
{\large
Vakhtang Kartvelishvili \footnote
{E-mail: vato@vxcern.cern.ch, vato@kheta.ge } and
Murman Margvelashvili \footnote
{E-mail: mmm@kheta.ge }}

\vspace{1.0cm}

{\em High Energy Physics Institute, Tbilisi State University,} \\
{\em GE-380086, Tbilisi, Georgia}

\vspace{0.3cm}

\end{center}
\vspace{1.0cm}

\begin{abstract}

The difference of $\tau$ partial decay widths into even and
odd number of pions, $R_{\tau,V-A}$, may be considered as
another important constraint on the
hadronic spectral function $\rho_{V-A}(s)$, in addition to four classic
sum rules of current algebra. Within certain reasonable
assumptions its value may be used to test the factorization
hypothesis and estimate the contribution
of dimension 6 condensates into the $\tau$ hadronic decay width.

\end{abstract}
\end{titlepage}
\renewcommand{\thefootnote}{\arabic{footnote}}
\setcounter{footnote}{0}
\vspace{0.4cm}
\newpage

{\bf 1.}
The difference of hadronic spectral densities with vector and axial vector
quantum numbers, $\rho_{V-A}(s)$, has been the object of special
attention since mid sixties \cite{WSR}, \cite{DMO}.
A number of theoretical
constraints imposed on this function have recently been
used for improving the
accuracy of vacuum condensate determination \cite{web} \cite{wef},
and even to "predict" the shape of the spectral function itself \cite{vk}.

In this note we recall that the difference of $\tau$ partial decay widths
into even and odd number of pions is another
directly measurable quantity which is expressed via the weighted integral
over the spectral function $\rho_{V-A}(s)$, and this relation may be
considered as an additional constraint over the hadronic spectra.
Provided the first Weinberg sum rule is valid and the
corrections coming from dimension 8 condensates are not unexpectedly large,
we are able toshow that the difference
of branching fractions can be used as the measure of the deviation
from vacuum saturation hypothesis, and
provides a way of estimating the contribution of the dimension 6
condensates into $\tau$ hadronic width. The latter is important in view
of the increased accuracies involved in tha $\alpha_s$ determination
from $\tau$ hadronic decay data.

\vspace{ 0.5 cm}

{\bf 2.} Consider the difference of $\tau$ decay widths
into even and odd number
of pions, normalized as usual to the leptonic decay width:
\begin{eqnarray}\label{rdef}
R_{\tau,V-A} & = & \sum_n{\frac{\Gamma(\tau\to\nu_{\tau}+2n\pi)-
\Gamma(\tau\to\nu_{\tau}+(2n+1)\pi)}
{\Gamma(\tau\to\nu_{\tau}+e+\nu_e)}} \nonumber \\
             & = & \sum_n{\frac{BR(\tau\to\nu_{\tau}+2n\pi)-
BR(\tau\to\nu_{\tau}+(2n+1)\pi)}
{BR(\tau\to\nu_{\tau}+e+\nu_e)}}.
\end {eqnarray}

In the following we will be working in the chiral limit, which is believed
to be a good approximation at the present level of accuracy, at least
as far as only $u$ and $d$ quarks are dealt with \cite{bra}, \cite{fnd};
thus we confine ourselves to the Cabibbo-allowed decays of the $\tau$.
In this case, the considered quantity can be expressed as
\begin{equation}\label{r}
R_{\tau,V-A}=12\pi\;|V_{ud}|^2\; \int_0^{M_\tau^2}{\frac{ds}{M_\tau^2}}
{(1-\frac{s}{M_\tau^2})^2}{(1+\frac{2s}{M_\tau^2})}
{\rm Im}\Pi_{V-A}(s+i\epsilon).
\end{equation}
Here $ \Pi_{V-A}(s)$ is the difference of vector and axial vector current
correlators, defined as
\begin{eqnarray}\label{pimn}
(-g^{\mu\nu}q^2 + q^\mu q^\nu) \Pi_{V-A}(q^2)&=&\\ \nonumber
 = i\int \,d^4x\,e^{-iqx}<0|&T&[V^\mu(x)V^\nu(0)^\dagger-
   A^\mu(x)A^\nu(0)^\dagger]|0>
\end {eqnarray}
with $V^\mu=\bar u \gamma^\mu d$ and $A^\mu=\bar u \gamma^\mu \gamma_5 d$,
while ${1\over{\pi}}{\rm Im}\Pi_{V-A}(s)=\rho_{V-A}(s)$ is the
difference of even- and odd-pion spectral densities, measured in $\tau$ decays.

Let us now remember that within SVZ approach \cite{svz} $\Pi_{V-A}$ has
the following theoretical expression:
\begin{equation}\label{ope}
\Pi_{V-A}(-Q^2)=-\frac{W_1}{Q^2}+\frac{W_2}{Q^4}+\frac{C^6<O^6>}{Q^6}+
\frac{C^8<O^8>}{Q^8}+\dots
\end {equation}
where by $W_1$ and $W_2$ we denote dimension 2 and 4 operators which
vanish in the chiral limit, while $C^6<O^6>$  and $C^8<O^8>$ come from
the four-quark operators of dimension 6 and 8. Note that pure
perturbative contributions cancel in the difference of vector and
axial vector operators (\ref{ope}).

Using the dispersion representation for $\Pi_{V-A}(Q^2)$ it is easy to obtain
a set of finite energy sum rules:
\begin{equation}\label{w1}
W_1=\int\rho_{V-A}(s)ds
\end {equation}
\begin{equation}\label{w2}
W_2=\int s\rho_{V-A}(s)ds
\end {equation}
\begin{equation}\label{c6}
C^6<O^6>=-\int s^2\rho_{V-A}(s)ds
\end {equation}
\begin{equation}\label{c8}
C^8<O^8>=\int s^3\rho_{V-A}(s)ds
\end {equation}

The first two of these equations coincide with the two Weinberg
sum rules \cite{WSR}
while the last two are FESRs used for determining the corresponding
condensates \cite{wef}.

One can easily notice that (\ref{r}) is essentially the combination of FESRs
(\ref{w1}),(\ref{c6}) and (\ref{c8}) with the upper limit
of integration substituted
by $M_{\tau}^2$. We believe however that this substitution is numerically
unimportant due to
the vanishing of $\rho_{V-A}(s)$ at high $s$ values and the double zero
of the integration weight at $s=M_{\tau}^2$. Comparing (\ref{w1}),(\ref{c6}),
(\ref{c8}) and (\ref{r}) we find
\begin{eqnarray}\label{rf1}
R_{\tau,V-A}&=&12\pi^2\;|V_{ud}|^2\;({\frac{W_1}{M_\tau^2}}
+\frac{3C^6<O^6>}{M_{\tau}^6}+\frac{2C^8<O^8>}{M_{\tau}^8})\\
\label{rf2}
&\approx&\frac{36\pi^2}{M_{\tau}^6}\,C^6<O^6>\,(1+\frac{2}{3M_{\tau}^2}
\frac{C^8<O^8>}{C^6<O^6>}),
\end {eqnarray}
where the validity of the first Weinberg sum rule $W_1=0$ was implied.
Note that the ratio $C^8<O^8>/C^6<O^6>$ is expected to be of order
 $(-1)\; GeV^2$ \cite{svz}, \cite{web}, \cite{wef}, so that the second term
in brackets in (\ref{rf2}) may be considered as a correction of order
$-(20\div 30)\%$. Hence, accurate measurement of (\ref{rdef}) would mean a good
estimate of the dimension 6 condensate, provided the spectral function
fulfils the first Weinberg sum rule.

\vspace{ 0.5 cm}

{\bf 3.}
Let us now try to extract some information about the size of dimension
6 condensate corrections to the {\em sum} of $\tau$ decay widths into
vector and axial vector Cabibbo allowed final states.
Following ref. \cite{bra} we will decompose $R_{\tau,V}$ and $R_{\tau,A}$
in the following way:
\begin{equation}\label{rdel}
R_{\tau,V/A}=\frac{3}{2}\;|V_{ud}|^2\;(1+\delta^0+\delta^2+
\delta^4+\delta^6_{V/A}+\delta^8_{V/A}),
\end{equation}
where $\delta^0$ stands for the perturbative correction, and $\delta^{n}$
 represent cotributions of dimension $n$ condensates. Note that
corrections originating from operators of dimension 2 and 4 which are
different for $V$ and $A$ channels, are expected to be small,
vanishing in the chiral limit \cite{bra}.

A commonly used parametrization for $\delta^6_{V/A}$
\cite{dosol} incorporates the
parameter $\rho$ which a measure of the deviation from the
vacuum saturation hypothesis:
\begin{equation}\label{cond}
\delta^6_{V/A}\simeq\left( \begin{array}{c} 7 \\ -11 \end{array} \right)
\frac{256\pi^3}{27}\frac{\rho \alpha_s <\bar{\psi}\psi>^2}{M_{\tau}^6}
\end{equation}
In this notation eq. (\ref{rf2}) can be rewritten as
\begin{equation}\label{rdel68}
R_{\tau,V-A}=\frac{3}{2}\;|V_{ud}|^2\;(\delta^6_{V-A}+\delta^8_{V-A})=
\frac{3}{2}\;|V_{ud}|^2\;\delta^6_{V-A}\;[1-(0.2\div 0.3)],
\end {equation}
where
\begin{equation}\label{d6va}
\delta^6_{V-A}\simeq\frac{512\pi^3}{3}
\frac{\rho \alpha_s <\bar{\psi}\psi>^2}{M_{\tau}^6},
\end {equation}
and an estimate of $\delta^8_{V-A}$ was substituted,
 as explained at the end of the previous section.
{}From eqs. (\ref{rdel68}-\ref{d6va}) one readily has:
\begin{equation}\label{rho}
\rho \alpha_s <\bar{\psi}\psi>^2\approx
\frac{M_{\tau}^6}{256\pi^3}\frac{1}{0.7\div 0.8}R_{\tau,V-A}=
(5.2\pm 0.5)\cdot10^{-3}\,GeV^6\cdot R_{\tau,V-A}.
\end{equation}

On the other hand, if one assumes the universality of the $\rho$ parameter
for different four-quark condensates, then
the dimension 6 correction to the total
hadronic width of the $\tau$, $\delta^6\equiv(\delta^6_{V}+\delta^6_{A})/2$,
can also be expressed through $R_{\tau,V-A}$ using eqs.
(\ref{cond}-\ref{d6va}):
\begin{eqnarray}\label{del6}
\delta^6 & \simeq &-\frac{512\pi^3}{27}
\frac{\rho \alpha_s <\bar{\psi}\psi>^2}{M_{\tau}^6} \nonumber \\
&\approx&(-\frac{2}{27})\frac{1}{0.7\div 0.8}\,R_{\tau,V-A}
\;\approx\;(-0.10\pm 0.01)\;R_{\tau,V-A}
\end {eqnarray}
Note that the errors quoted in (\ref{rho}) and (\ref{del6}) account
only for the uncertainty in $\delta^8_{V-A}$, which was assumed to be
around $\pm30\div40\%$. This shows that eqs. (\ref{rho}) and (\ref{del6})
are not too sensitive to the above assumption.

\vspace{ 0.5 cm}

{\bf 4.}
Recently ALEPH collaboration has published a complete set of measured
branching ratios of $\tau$ decays into $h$, $h\pi^0$, $h2\pi^0$,
$3h$, $h3\pi^0$, $3h\pi^0$ final states \cite{al}. Naive
substitution into (\ref{rdef}) with summing up the errors in
quadrature yields:
\begin{equation}\label{alexp}
R^{exp}_{\tau,V-A}=0.022\pm 0.083.
\end{equation}
Eqs. (\ref{rho}) and (\ref{del6}) then give:
\begin{equation}\label{rho1}
\rho \alpha_s <\bar{\psi}\psi>^2=(1.1\pm4.0)\cdot 10^{-4} \, GeV^6,
\end{equation}
\begin{equation}\label{del61}
\delta^6=(-2.2\pm8.3)\cdot10^{-3},
\end {equation}
which should be compared to the values used in the analysis \cite{bra}:
\begin{equation}\label{rho2}
\rho \alpha_s <\bar{\psi}\psi>^2=(3.8\pm2.0)\cdot 10^{-4} \, GeV^6,
\end{equation}
\begin{equation}\label{del62}
\delta^6=(-7\pm4)\cdot10^{-3},
\end {equation}
We must admit that the experimental value (\ref{alexp}) should not
be taken too seriously for two reasons: first, no charged $\pi-K$
separation was attempted in the experiment \cite{al};
secondly, when summing up the errors in quadrature we have neglected
obviously strong correlations between various decay rates.
Still, we believe that future measurements will improve the situation.
With the substantial increase in statistics one may expect dramatic
improvements in the accuracies of the measured branching fractions of
the $\tau$ hadronic decays, and a dedicated study of the {\em difference}
of branching ratios (\ref{rdef}) might result in the error in
$R_{\tau,V-A}$ as good as 0.01, thus making our formulas
(\ref{rho}) and (\ref{del6}) quite useful.

At the same time it would be highly important to measure accurately
the spectral function $\rho_{V-A}(s)$,  check the validity
of Weinberg sum rules, and estimate dimension 6 and 8 condensates
using some other techniques like Finite Energy or Borel-transformed
sum rules. All this could be a good self-consistency check for both
theory and experiment.

\end{document}